\documentclass[aps,pra,preprint,superscriptaddress]{revtex4-2}
\usepackage{graphicx}
\usepackage{float}
\usepackage{color}
\usepackage{amsmath}
\usepackage[colorlinks=true, citecolor=red, linkcolor=blue,urlcolor=blue ]{hyperref}
\usepackage{appendix}
\usepackage{mathtools}
\usepackage{booktabs}
\usepackage{txfonts}
\usepackage{layouts}
\usepackage{verbatim}
\usepackage{upgreek}
\usepackage{pgffor,xprintlen}
\usepackage{orcidlink}

\newcommand{\um}{$\mathrm{\upmu m}$}
\newcommand{\Afn}{\overline{A}_{f\parallel}(N)}
\newcommand{\Bfn}{\overline{B}_f(N)}
\newcommand{\Cfn}{\overline{A}_{f\perp}(N)}
\newcommand{\Dfn}{\overline{D}_f(N)}
\newcommand{\ez}{\mathbf{e}_z}
\newcommand{\sign}{\text{sgn}}

\begin{document}
\title{Physical mechanism reveals bacterial slowdown above a critical number of flagella}
	
\author{Maria T\u{a}tulea-Codrean\,\orcidlink{0000-0001-5518-7958}}
\email[]{m.tatulea-codrean@damtp.cam.ac.uk}
\affiliation{Department of Applied Mathematics and Theoretical Physics, University of Cambridge, Cambridge CB3 0WA, United Kingdom}

\author{Eric Lauga\,\orcidlink{0000-0002-8916-2545}}
\email[]{e.lauga@damtp.cam.ac.uk}
\affiliation{Department of Applied Mathematics and Theoretical Physics, University of Cambridge, Cambridge CB3 0WA, United Kingdom}
	
\date{\today}
	
\begin{abstract}
    Numerous studies have explored the link between bacterial swimming and the number of flagella, a distinguishing feature of motile multiflagellated bacteria. We revisit this open question using augmented slender-body theory simulations, in which we resolve the full hydrodynamic interactions within a bundle of helical filaments rotating and translating in synchrony. Unlike previous studies, our model considers the full torque-speed relationship of the bacterial flagellar motor, revealing its significant impact on multiflagellated swimming. Because the viscous load per motor decreases with flagellar number, the {bacterial flagellar motor (BFM)} transitions from the high-load to the low-load regime at a critical number of filaments, leading to bacterial slowdown as further flagella are added to the bundle. We explain the physical mechanism behind the observed slowdown as an interplay between the load-dependent generation of torque by the motor, and the load-reducing cooperativity between flagella, which consists of both hydrodynamic and non-hydrodynamic components. The theoretically predicted critical number of flagella is remarkably close to the values reported for the model organism \textit{Escherichia coli}. Our model further predicts that the critical number of flagella increases with viscosity, suggesting that bacteria can enhance their swimming capacity by growing more flagella in  more viscous environments, consistent with empirical observations.		
\end{abstract} 
	
\maketitle 

%Column width is \printlen[2][cm]{\columnwidth}
%Text width is \printlen[2][cm]{\textwidth}
%Line width is \printlen[2][cm]{\linewidth}

\section{Introduction}

Bacterial flagella are filamentous appendages that enable bacteria to swim \cite{Berg1973,Berg2004}. Each flagellum consists of a rigid helical filament and a rotary motor, connected by a flexible joint known as the hook \cite{Armitage2020}. The physics of flagellar motility relies on the combined effect of the rotary actuation provided by the flagellar motor, the chiral structure of the flagellar filament, and the anisotropic viscous drag exerted on filaments by the surrounding fluid \cite{Purcell1977}. These basic hydrodynamic principles of locomotion apply to all flagellated bacteria from monotrichous (single flagellum) to peritrichous bacteria (multiple flagella) \cite{Lauga2015}.

The function of bacterial multiflagellarity remains elusive despite extensive theoretical and experimental work on this subject \cite{Kanehl2014,Nguyen2018,Najafi2018,Kamdar2023}. Unlike some unicellular algae which have a fixed number of cilia synchronized in well-defined swimming gaits \cite{Wan2016}, multiflagellated bacteria are equipped with a variable number of flagella either scattered around the cell body (peritrichous flagellation) or clustered at one pole (polar flagellation) \cite{leifson1960atlas}. The model organism for peritrichous bacteria, \textit{Escherichia coli} (\textit{E.~coli}), typically has between two and five flagella \cite{Turner2000,Turner2016}. While the bacterial flagellum has multiple functions \cite{Chaban2015}, the focus of this study will be on its primary role as a locomotory organelle. 

In the context of locomotion, multiflagellarity is an ingenious navigation strategy as it allows bacteria to explore complex environments by modulating the frequency of `runs' and `tumbles' in response to chemical cues \cite{Berg1972,Berg1983}. During a `run', bacteria swim in a straight line with their flagella bundled at the back of the cell body. The bundles are periodically disrupted by motor reversals from counter-clockwise to clockwise rotation leading to flagellar unbundling and a brief period of reorientation known as a `tumble'. Since the reversal of any single motor can induce a tumbling event, the duration of straight runs is expected to decrease with the number of flagella, in line with recent observations \cite{Najafi2018}. Furthermore, having multiple flagella may stabilize the swimming direction and make runs more effective \cite{Nguyen2018}. In contrast to this stabilizing effect during run periods, a theoretical model of tumbling has predicted that the flagellar number correlates positively with the mean tumbling angle (i.e.~the angle between consecutive runs) due to a combination of hydrodynamic and geometrical factors \cite{Dvoriashyna2021}. This is qualitatively consistent with experimental observations \cite{Najafi2018}.

Here, we address a long-standing question about bacterial multiflagellarity: how does the swimming speed of bacteria depend on the number of flagella, and what sets the flagellar numbers consistently observed within some multiflagellated species? Previous computational studies have predicted that the swimming speed of multiflagellated bacteria grows sub-linearly with the number of flagella, but have not identified any motility advantage that would favour the selection of a specific number of flagella \cite{Kanehl2014,Nguyen2018}. Experimentally, it is more difficult to isolate the number of flagella as a control parameter since it correlates positively with cell body size. {In a genetically identical population of \textit{E. coli} bacteria, it has been found that natural variations in flagellar number and cell body size result in a constant swimming speed across a wide range of bacterial lengths, due to the balance between the higher propulsive force generated by an increasing number of flagella and the larger hydrodynamic drag on bacteria with a longer cell body} \cite{Kamdar2023}. Genetic manipulations have been performed on \textit{Bacillus subtilis}---to create strains with different flagellar numbers---and have indicated that the number of flagella affects several motility markers including the speed, tumbling frequency, and tumbling angle of bacteria, making hypo-flagellated mutants better adapted for long-range transport and hyper-flagellated mutants better suited to biofilm formation \cite{Najafi2018}. In this study, we focus on the swimming speed of bacteria during straight `runs' when the flagella rotate together at the back of the cell body in a parallel bundle. Computationally, we are able to vary the number of flagella while keeping all other parameters constant, thereby isolating the parameter of interest for this investigation. 

\begin{figure}[t!]
    \centering
    \includegraphics[width=.95\textwidth]{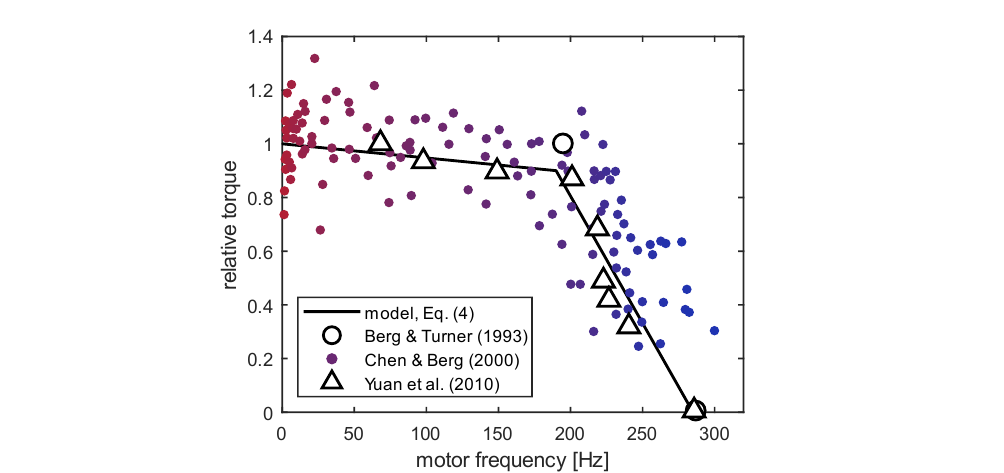}
    \caption{\textbf{Data-driven model for torque generation in the bacterial flagellar motor (BFM)}. The torque-speed relationship of the BFM is modelled as a piecewise linear curve (black) in agreement with experimental measurements of the \textit{Escherichia coli} (\textit{E.~coli}) motor at room temperature \cite{Berg1993,Chen2000,Yuan2010}. Filled circles indicate individual measurements, while hollow circles and triangles represent binned averages. {The torque is normalised by the `stall torque' of the motor at zero frequency, for which we select the value in Table \ref{table}}.}
    \label{fig1}
\end{figure}

What our study demonstrates is the key role played by the bacterial flagellar motor (BFM) \cite{Sowa2008,Nirody2017}. While previous computational studies assumed that the motor operates at constant torque \cite{Kanehl2014,Nguyen2018}, we aim to characterize the dynamics of the BFM accurately across a wide range of hydrodynamic loads. When a large hydrodynamic load is attached to the motor, such as a long flagellum or a large polystyrene bead \cite{Sowa2003,Nord2022}, the motor operates at nearly-constant torque (`high-load' regime). However, when a small hydrodynamic load such as a flagellar stub is in place, the rotation speed is nearly constant (`low-load' regime). Hence, the resulting torque-speed curve features a kink or `knee' between the high-load and low-load regimes. The torque-generating capacity of the BFM is represented schematically in Fig.~\ref{fig1} as a piecewise linear curve, { alongside experimental measurements 
\footnote{Data points were manually extracted from Fig.~16 \cite{Berg1993}, Fig.~5a \cite{Chen2000}, and Fig.~1 \cite{Yuan2010} using a web-based data extraction tool \cite{WebPlotDigitizer}\color{black}}.}
{The torques are scaled by the `stall torque' of the motor, which serves as a model parameter and is set by default to the value listed in Table \ref{table}.}

In this article, we present a new computational investigation into the dependence of bacterial swimming speed on the number of flagella. Our kinematic model of a swimming bacterium consists of a spheroidal cell body and a bundle of regularly spaced helical filaments rotating in parallel and in phase with each other (Fig.~\ref{fig2}a). The filaments are fully coupled through hydrodynamic interactions (HIs) resolved via an augmented slender-body theory (SBT+) \cite{Tatulea-Codrean2021,Tatulea-Codrean2022,Kamdar2023}, while the flagellar bundle and the cell body are coupled through a global force and torque balance (Eqs.~\eqref{eq:force-balance}-\eqref{eq:torque-balance}, dry coupling). % through the force and torque balance on the entire cell
Crucially, our model takes into account the full torque-speed relationship of the BFM as an additional constraint (Eqs.~\eqref{eq:local-torque-balance}-\eqref{eq:BFM-torque-speed}, Fig.~\ref{fig2}b), in contrast to previous studies which assumed that the BFM operates at a constant torque. Thus, we find that the motors transition from high-load to low-load regime as the number of filaments increases due to a reduction in the viscous load per filament (Fig.~\ref{fig2}b,c). This leads to a previously unreported slowdown in bacterial swimming above a critical number of flagella, meaning that the bacterium would incur a penalty in motility if it assembled too many flagella (Fig.~\ref{fig2}d,e). Remarkably, the theoretically predicted critical number is very close to the flagellar numbers reported for \textit{E.~coli} in the literature \cite{Turner2000,Turner2016}. We further demonstrate that { the physical mechanism for bacterial slowdown consists in a load-dependent torque generating capacity (BFM torque-speed relation) coupled with a load-reducing cooperativity between propellers (hydrodynamic or otherwise)} (Fig.~\ref{fig3}). { A non-hydrodynamic source of cooperativity between the flagella is the mechanism of `cargo' load sharing, which we illustrate using an electrical analogy (Fig.~\ref{fig4})}. Finally, our theory predicts that bacteria could maximize their swimming speed by growing more flagella in media of higher viscosity, a verifiable prediction which is consistent with empirical observations (Fig.~\ref{fig5}). {We conclude the study by examining the sensitivity of the speed-flagellar number relationship to changes in the motor parameters, in order to test the applicability of our results to experimental data from studies of other bacterial species with distinct torque-speed curves (Fig.~\ref{fig6}).}

\section{Modelling the swimming of a multiflagellated bacterium}

\begin{figure}[p!]
	\includegraphics[width=\textwidth]{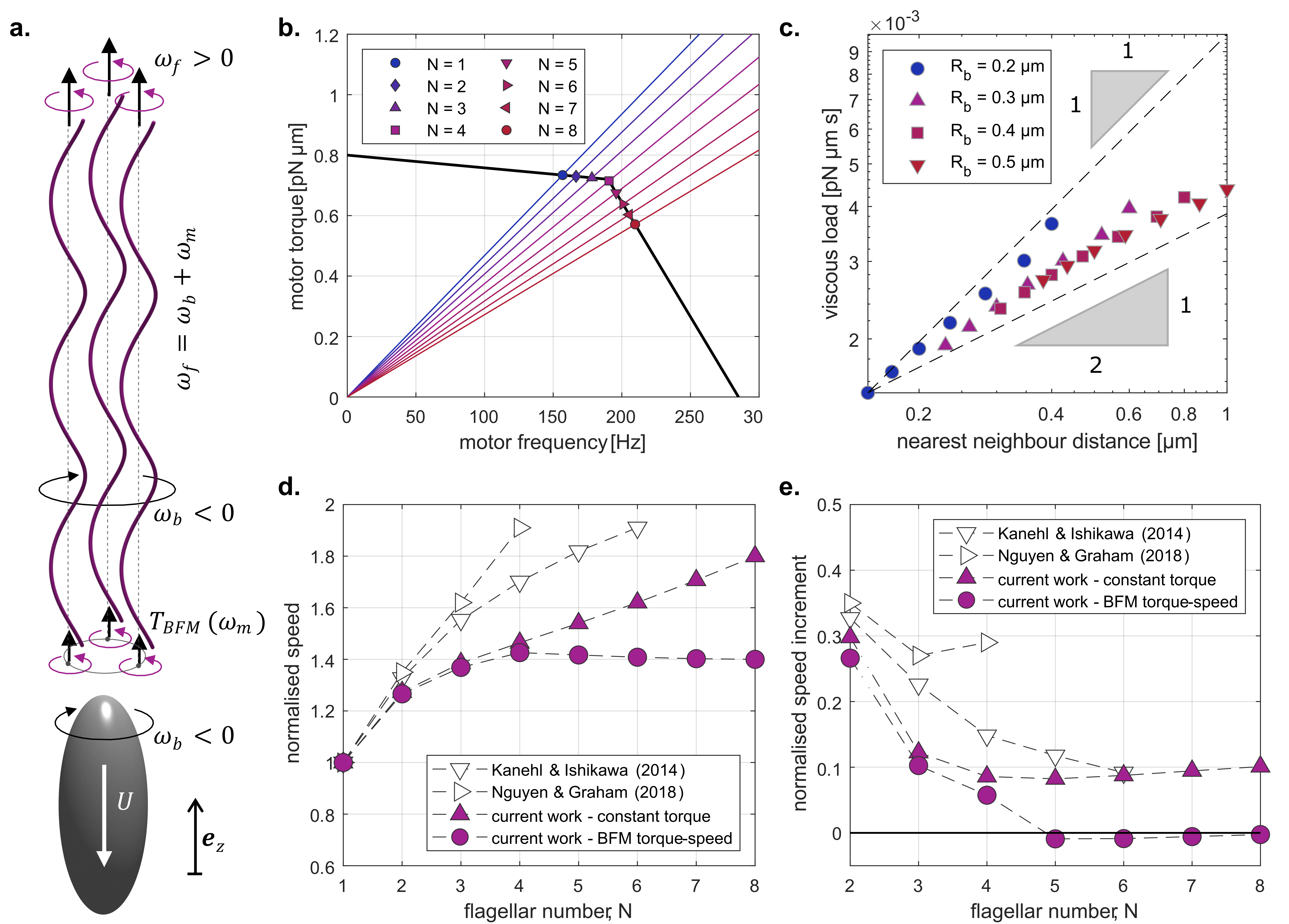}
	\caption{\textbf{Kinematic model of bacterial swimming and main results}. (a) The flagellar filaments are phase-locked and rotate ($\omega_f$) around prescribed axes of rotation. The flagellar bundle and cell body rotate ($\omega_b$) and translate ($U$) together along the major axis of the body. (b) The motor operates at the intersection of its torque-generating capacity (black curve) and the characteristic load lines of the flagellar filament (colour lines). The `load' on each motor (slope of load line) decreases with increasing number of filaments ($N$) in the bundle. (c) The viscous load per motor depends on both the number of filaments and the radius of the flagellar bundle, with a strong positive correlation between the load and the nearest-neighbour distance between filaments. (d,e) When the motors operate at constant torque (triangles), the swimming speed increases monotonically with flagellar number (d), and each additional filament generates a positive speed increment (e). However, when the motors obey the characteristic BFM torque-speed relationship (circles), the bacterium is predicted to slow down above a critical number of flagella (d), due to negative speed increments (e). Both the speed, $U(N)$, and speed increment, $U(N)-U(N-1)$, are normalised by the swimming speed for a single flagellum, $U(1)$.}
	\label{fig2}
\end{figure}

As shown in Fig.~\ref{fig2}a, we model the swimming multiflagellated bacterium as a rigid prolate spheroid (the cell body) and $N$ identical rigid helical filaments (the flagella) moving together with velocity $U\ez$ (here, $U<0$). Looking down on the flagellar bundle and towards the cell body, a stationary observer would see the filaments rotate counter-clockwise and the cell swim downwards and away from them (see Fig.~\ref{fig2}a). In the laboratory frame, the flagellar filaments rotate with angular velocity $\omega_f\ez$ (here, $\omega_f >0$), while the cell body rotates in the opposite direction with angular velocity $\omega_b\ez$ (here, $\omega_b <0$) to balance the torque, such that the flagellar motors embedded in the cell wall rotate with speed $\omega_m = \omega_f-\omega_b$. In our model, the filaments rotate in phase with each other around their individual axes of rotation, which are regularly spaced around a cylinder of radius $R_b$, the radius of the flagellar bundle. The entire bundle also rotates clockwise around the major axis of the cell with angular velocity $\omega_b\ez$ together with the cell body. Our chosen parameter values for the cell body and filaments, listed in Table \ref{table}, correspond to a typical \textit{E. coli} bacterium with flagella in the `normal' polymorphic form \cite{Turner2000,Darnton2007,Turner2016}. 

\subsection{Force and torque balance}

The cell body and flagellar bundle are coupled through the Stokes flow constraint of zero net force and torque on the entire cell \cite{Purcell1977}, imposing that
\begin{eqnarray}
A_b U + N\left(\Afn U + \Bfn \omega_f \right) &=& 0, \label{eq:force-balance}\\
\left(D_b+NR_b^2\Cfn\right) \omega_b + N\left(\Bfn U + \Dfn \omega_f\right) &=& 0, \label{eq:torque-balance}
\end{eqnarray}
where $\Afn$ is the effective drag on a filament translating parallel to its axis of rotation. Similarly, $\Bfn$ and $\Dfn$ are the effective thrust and torque per unit angular velocity of a rotating filament. Due to the symmetry of the hydrodynamic resistance matrix, the coefficient $\Bfn$ represents both the force per unit angular velocity of a rotating filament, and the torque per unit linear velocity of a translating filament. The drag coefficient $\Cfn$ represents the effective resistance of each filament to motion perpendicular to its axis of rotation. Because the bundle co-rotates with the cell body at angular velocity $\omega_b$, and the filaments are placed at a distance $R_b$ away from the axis of rotation of the cell body, this gives rise to a torque proportional to the square of the bundle radius. The explicit dependence of the effective drag coefficients on flagellar number indicates that each filament belongs to a bundle of $N$ hydrodynamically interacting filaments with identical geometry and kinematics. There is an implicit dependence on other parameters in the system including the placement, helical geometry, and contour length of the flagellar filaments.

\subsection{BFM torque-speed relationship}

To close the system of equations, we impose that the net hydrodynamic torque on each flagellar filament matches the torque generated by the BFM, 
\begin{equation}
\Bfn U + \Dfn \omega_f = T_{BFM}(\omega_f - \omega_b),
\label{eq:local-torque-balance}
\end{equation}
and we model the BFM torque-speed relationship as a piecewise linear function of the motor angular velocity,
\begin{equation}
    T_{BFM}(\omega_m) = T_{knee} - \left(\omega_m-\omega_{knee}\right)\alpha(\omega_m-\omega_{knee}), 
    \label{eq:BFM-torque-speed}
\end{equation}
where the linear slope $\alpha$, given by the functional form $\alpha(\xi) = \overline{\alpha} + \Delta\alpha~\sign(\xi)$, takes distinct values in the high-load ($\omega_m<\omega_{knee}$) and low-load ($\omega_m>\omega_{knee}$) regimes. { For the knee and zero-torque frequencies of the motor, as well as the relative knee torque, we use the values reported by \citet{Yuan2010}, which are in good agreement with previous observations of the \textit{E.~coli} flagellar motor at room temperature {(Fig.~\ref{fig1})}. Our chosen value for the stall torque is within the range computed by \citet{Das2018}. These values, together with all other dynamic and geometric parameters, are listed in Table \ref{table}.}

\begin{table}[ht!]
\centering
\begin{tabular}{l l l}
\toprule
Parameter ~~~~~~~~~~~~~~~~~~~~~~~~~~~~~~~~~~~~~~~~~~~ & Symbol ~~~~~ & Value \\
\midrule
Cell body, length & $l$ & $2.5$ \um \\
Cell body, width & $w$ & $1.0$ \um \\
Filament, contour length & $L$ & $8.0$ \um \\
Filament, helical pitch & $p$ & $2.5$ \um \\
Filament, helical radius & $r$ & $0.21$ \um \\
Filament, cross-sectional radius & $r_\epsilon$ & $12$ nm \\
Bundle, radius & $R_b$ & $0.5$ \um \\
Dynamic viscosity & $\mu$ & $0.93$ cP \\
Motor, stall torque & $T_{\mathrm{stall}}$ & {$0.80$ pN \um} \\
Motor, knee torque & $T_{\mathrm{knee}}$ & ${0.9}\times T_{stall}$\\
Motor, knee frequency & $f_{\mathrm{knee}}$ & {$190$ Hz} \\
Motor, zero-torque frequency & $f_0$ & {$285$ Hz} \\
\midrule
Motor, knee angular velocity & $\omega_{\mathrm{knee}}$ & $1.2\times 10^3$ rad/s \\
Motor, high-load slope & $\overline{\alpha} - \Delta\alpha$ & $6.7\times 10^{-5}$ pN \um~s/rad\\
Motor, low-load slope & $\overline{\alpha} + \Delta\alpha$ & $1.2\times 10^{-3}$ pN \um~s/rad\\
\bottomrule 
\end{tabular}
\vspace{0.4cm}
\caption{\textbf{Default parameter values}. The filament geometry corresponds to the `normal' polymorphic form of bacterial flagella \cite{Turner2000,Darnton2007,Turner2016}, the viscosity is that of water at 23$^{\circ}$C, and the motor parameters characterize the torque-speed relationship of the \textit{E.~coli} proton-driven motor at room temperature \cite{Berg1993,Chen2000,Yuan2010}. {The stall torque falls within the range computed by \citet{Das2018}.} In the present study, each parameter assumes the value listed above, unless that parameter is varied on the abscissa or in the legend of a given figure. The last three parameters are calculated according to $\omega_{\mathrm{knee}} = 2\pi f_{\mathrm{knee}}$, $\overline{\alpha} - \Delta\alpha = (T_{\mathrm{stall}}-T_{\mathrm{knee}})/\omega_{\mathrm{knee}}$, $\overline{\alpha} + \Delta\alpha = T_{\mathrm{knee}}/(\omega_0 - \omega_{\mathrm{knee}})$.}
\label{table}
\end{table}

\subsection{Solution for swimming speed and effective load on motor}

The governing equations, Eqs.~\eqref{eq:force-balance}-\eqref{eq:BFM-torque-speed}, constitute a linear system of equations in three unknowns: the swimming speed of the cell, $U$, the angular velocity of the flagella, $\omega_f$, and the angular velocity of the cell body, $\omega_b$. We invert this system analytically to find that the swimming speed is
\begin{equation}
    U = -\frac{\lambda\omega_m}{2\pi}, \label{eq:U}
\end{equation}
and each motor rotates with angular velocity
\begin{equation}
    \omega_m = \frac{\delta_{knee} + \alpha(\delta_{knee}-\delta)}{\delta + 
    \alpha(\delta_{knee}-\delta)}\omega_{knee}, \label{eq:Om_m}
\end{equation}
where we have introduced the step length, $\lambda$, and the viscous load, $\delta$, defined as
\begin{eqnarray}
    \lambda &=& 2\pi \Bfn\frac{\frac{D_b}{N} + R_b^2\Cfn}{\left(\frac{A_b}{N} + \Afn\right)\left(\frac{D_b}{N} + R_b^2\Cfn+\Dfn\right) -\Bfn^2}, \label{eq:lambda-steplength}\\
    \delta &=& \left[\left(\frac{A_b}{N} + \Afn\right)\Dfn-\Bfn^2\right]\frac{\frac{D_b}{N} + R_b^2\Cfn}{\left(\frac{A_b}{N} + \Afn\right)\left(\frac{D_b}{N} + R_b^2\Cfn+\Dfn\right) -\Bfn^2}. \label{eq:delta-load}
\end{eqnarray}
The step length, $\lambda$, represents the distance travelled by the swimming bacterium in one complete revolution of the motor, see Eq.~\eqref{eq:U}, while the parameter $\delta$ represents the effective viscous load on a single flagellar motor, thereby combining the hydrodynamic resistance to rotation of the flagellar filaments and the cell body. The effective viscous load sets the slope of the `load lines' in Fig.~\ref{fig2}b. When the motor operates in the high-load (or low-load) regime, the effective viscous load is higher (or lower) than the viscous load at the knee, $\delta_{knee} = T_{knee}/\omega_{knee}$.

\subsection{Hydrodynamic modelling of cell body and flagella}

The derivation of Eqs.~\eqref{eq:U}-\eqref{eq:delta-load} is independent of the choice of hydrodynamic model for the drag coefficients of the cell body and the flagella. Here, we model the cell body as a prolate spheroid of length $l$, width $w$, and eccentricity $e = \sqrt{1-w^2/l^2}$. The exact drag coefficients for translation $A_b = 8\pi\mu le^3\left[-2e+(1+e^2)\ln{\left(\frac{1+e}{1-e}\right)}\right]^{-1}$, and rotation $D_b = 4\pi\mu l^3e^3(1-e^2)\left[3\left(2e+(1-e^2)\ln{\left(\frac{1+e}{1-e}\right)}\right)\right]^{-1}$ parallel to the major axis of the spheroid have been calculated analytically \cite{Chwang1975}.

For the filaments, the effective drag coefficients $\Afn, \Cfn, \Bfn, \Dfn$ are computed numerically and averaged over the phase angle, $\phi$, of an individual filament within an equally spaced bundle of phase-locked rotating filaments, as shown in Fig.~\ref{fig2}a. Specifically, we have $\Afn = \frac{1}{2\pi}\int_0^{2\pi} A_{f\parallel}(N,\phi)\mathrm{d}\phi$. For each configuration, the instantaneous drag coefficients $A_{f\parallel}(N,\phi)$, $A_{f\perp}(N,\phi)$, $B_{f}(N,\phi)$, $D_{f}(N,\phi)$ are computed using a custom-build MATLAB implementation of Johnson's slender-body theory (SBT) \cite{Johnson1980} enhanced with hydrodynamic interactions for the case of multiple filaments \cite{Tornberg2004}. We will refer to this method as augmented slender-body theory (SBT+). In our implementation of SBT+, the force densities and velocity distributions along the centerlines of the filaments are projected onto a convenient basis of Legendre polynomials \cite{Gotz} { and truncated to a suitable number of modes ($N_{\text{Legendre}}=15$ in this study). Our implementation of SBT+ was previously validated and described in full detail in our earlier studies of bacterial flagellar dynamics \cite{Tatulea-Codrean2021,Tatulea-Codrean2022,Kamdar2023}.

Solving for the HIs between filaments in the linear Stokes flow regime becomes a linear inverse problem between the Legendre coefficients of a prescribed set of filament velocities and the Legendre coefficients of an unknown set of force densities. The coefficients for different filaments are coupled to each other through the long-range flows induced by each moving filament \cite{Tornberg2004}. The final computational output of our SBT+ implementation is an extended hydrodynamic resistance matrix which describes the force and torque on each given filament due to the rigid-body motion of any other, from which we extract the instantaneous drag coefficients $A_{f\parallel}(N,\phi)$, $A_{f\perp}(N,\phi)$, $B_{f}(N,\phi)$, $D_{f}(N,\phi)$. To reduce computational costs, we only simulate the rotating filaments over the interval $\phi\in(0,2\pi/N)$ and extend our results to a full period of rotation $\phi\in(0,2\pi)$ using the rotational symmetry in the bundle}. 

\section{Theoretical results on multiflagellated swimming}

We now use the kinematic model described in the previous section to gain new insights into the swimming of multiflagellated bacteria. First, we focus on the effect of introducing the BFM torque-speed relationship into our model and find that the swimming speed no longer grows monotonically with  the number of flagella (Fig.~\ref{fig2}). Bacteria with a fixed body size are expected to slow down if the number of flagella increases above a critical number. Next, we investigate the physical mechanism responsible for this observation and conclude that it requires a combination of (i) load-dependent torque generation by the motor, and (ii) load-reducing cooperativity between the flagella, which need not be of hydrodynamic origin (Fig.~\ref{fig3}). We also illustrate the concept of non-hydrodynamic load-sharing between motors using an electrical analogy (Fig.~\ref{fig4}). Finally, we explore the effect of changing the environmental viscosity on the critical number of flagella and draw new insights about the mechanical role of multiflagellarity (Fig.~\ref{fig5}).

\subsection{BFM transitions between high-load and low-load regime as number of flagella increases}

First, we observe that the viscous load per motor decreases with the number of flagella, as shown by the fan of load lines {emanating from the origin} in Fig.~\ref{fig2}b. Our theory predicts that the BFM transitions between the high-load and low-load regimes at a biologically relevant number of flagella: {around four flagella} in Fig.~\ref{fig2}b. This highlights the importance of including the BFM torque-speed relationship into models of bacterial swimming, since the BFM deviates significantly from the regime of nearly-constant torque under realistic swimming conditions.

We also find that the viscous load per motor depends on both the number of flagella and the radius of the bundle, $R_b$. When plotted against the distance between two nearest neighbours in the flagellar bundle, $d_{NN} = 2R_b \sin (\pi/N)$, the viscous load collapses onto a power law with exponent between 0.5 and 1 (Fig.~\ref{fig2}c). Because the transition from high-load to low-load regime is controlled by the balance between the effective load per motor (set by the geometry of the flagellar filaments and of the cell body) and the knee load (set by the characteristics of the motor), all the model parameters listed in Table \ref{table} will have an impact on this transition and the predictions that follow from it. 

\subsection{Torque reduction in BFM low-load regime leads to bacterial slowdown above critical number of flagella}

As the viscous load drops, the BFM cannot provide as much torque to the flagella, which raises an obvious question: how does the reduction in motor torque affect the swimming speed of the bacterium? We plot the swimming speed as a function of flagellar number in Fig.~\ref{fig2}d, comparing our theoretical predictions with those previously reported in the literature by \citet{Kanehl2014} and \citet{Nguyen2018}. For the latter study, we report the numerical results on so-called `polar' bacteria because this flagellar placement most closely resembles that in the current study. At constant motor torque, both our kinematic model and previous elastohydrodynamic models predict that the swimming speed monotonically increases with the number of flagella (triangles, Fig.~\ref{fig2}d). However, when we impose the correct BFM torque-speed relationship, our theory predicts that the swimming speed {plateaus} at a finite number of flagella (circles, Fig.~\ref{fig2}d). This is due to a reduction in the torque supplied to each propeller after the BFM transitions to the low-load regime. To quantify the motility benefit of adding another flagellum to the bundle, we also plot the speed increment, $U(N)-U(N-1)$, against the number of flagella (Fig.~\ref{fig2}e). The bacterial slowdown predicted by our kinematic model is associated with {small} negative speed increments above a `critical number of flagella', which is defined as the greatest integer associated with a positive speed increment.

\subsection{Physical mechanism for bacterial slowdown combines the load-dependent generation of torque by motors with the load-reducing cooperativity between flagella}

\begin{figure}[t!]
	\centering
        \includegraphics[width=.9\textwidth]{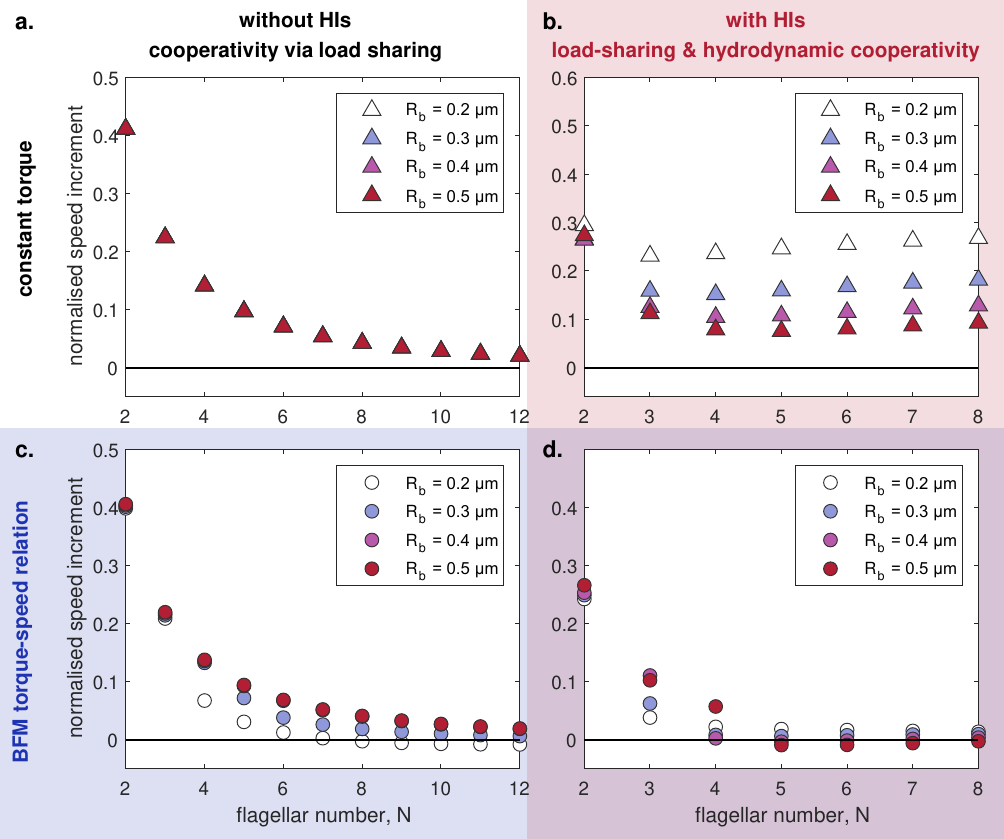}
	\caption{\textbf{Physical mechanism for bacterial slowdown}. By systematically introducing into our computational model the HIs between flagellar filaments (rightmost panels) and the torque-speed relationship of the BFM (lower panels), we are able to confirm that the observed slowdown is due primarily to the torque-generating capacity of the BFM (c). On their own, HIs cannot generate negative speed increments (b) but they do contribute to bringing the critical flagellar number within a biologically relevant range (d). The speed increments, $U(N)-U(N-1)$, are normalised by the swimming speed of a monoflagellated bacterium, $U(1)$.}
	\label{fig3}
\end{figure}

What is the underlying cause of the slowdown observed in the previous subsection? To establish the physical mechanism behind this phenomenon, we systematically introduce into our model the HIs between flagellar filaments, and the BFM torque-speed relationship. In Fig.~\ref{fig3}, we show the speed increments due to the addition of the $N$th flagellar filament, $U(N)-U(N-1)$, and we compare the results obtained using different modelling assumptions: with (b,d) or without (a,c) HIs between the filaments, and with (c,d) or without (a,b) the BFM torque-speed relationship. Note that the results shown in Figs.~\ref{fig3}a,c are accessible to the general reader via the analytical results from Eqs.~\eqref{eq:U}-\eqref{eq:delta-load}, as no computational step is required for the HIs between filaments. It is possible to estimate the drag coefficients $A_{f\parallel}$, $A_{f\perp}$, $B_{f}$, $D_{f}$ of a helical filament analytically using resistive-force theory instead of SBT \cite{Tatulea-Codrean2021}.

First, we observe that negative speed increments only emerge when the BFM operates according to its characteristic torque-speed curve (Figs.~\ref{fig3}c, d). This demonstrates that the load-dependent generation of torque is an essential physical ingredient for the observed slowdown. The only way for this mechanism to manifest as the number of flagella increases is if the load on each motor is affected by the number of flagella. For bacteria to slow down, the torque supplied by each motor would need to decrease, suggesting that a load-reducing cooperativity between flagella is also required. 

One source of cooperativity are the HIs between flagellar filaments. It is known that helical filaments rotating in parallel require less actuating torque than filaments rotating on their own at the same angular velocity (see Fig.~7f in Ref.~\cite{Tatulea-Codrean2021}). Suppose a left-handed helical filament, corresponding to the `normal' polymorphic shape of bacterial flagella \cite{Spagnolie2011}, is rotating counter-clockwise with a fixed angular velocity $+\mathbf{e}_z$ (this defines the direction of the $z$-axis). On average, this filament would be pumping fluid along its axis of rotation in the positive $z$ direction. A neighbouring identical filament would respond to this upward flow by rotating counter-clockwise, even in the absence of external forces or torques on the second filament. This is because of the anisotropic Stokes drag on slender filaments, which dictates that the local force density induced by the upward flow on the left-handed helical filament has a component in the positive $\phi$ direction. These contributions add up to a net counter-clockwise torque induced by the first filament on the second. If this torque is not resisted by any external mechanism, the second filament will rotate counter-clockwise. Since helical filaments rotate more easily together than on their own, these additive constructive HIs contribute to the decrease in the effective load per motor as the number of flagella increases.

Surprisingly, the slowdown occurs even in the absence of HIs (Fig.~\ref{fig3}c), suggesting that the load-reducing cooperativity between flagella goes beyond hydrodynamic effects, as we explain in the next subsection. Nevertheless, HIs are important because they bring the critical number of flagella within the biologically relevant range for \textit{E.~coli} bacteria (from {seven} or more flagella in Fig.~\ref{fig3}c, to three or four flagella in Fig.~\ref{fig3}d).

In summary, the physical mechanism for bacterial slowdown combines the load dependence of torque generation by the flagellar motors (kinked torque-speed curve in Fig.~\ref{fig2}b, in black) and the load reduction due to cooperativity between flagella (fan of load lines in Fig.~\ref{fig2}b, in colour).

\subsection{Non-hydrodynamic cooperativity between flagella via cargo load sharing}

To understand the non-hydrodynamic component of load reduction, we need to make a distinction between the load or hydrodynamic resistance to rotation of the cell body (the `cargo') and that of the filaments (the `propellers'). The effective load per motor decreases with the number of flagella---even without HIs---because each motor is responsible for rotating one propeller but all motors share the load of the cargo. 

While this fact is not immediately obvious from Eq.~\eqref{eq:delta-load}, the expression for the effective load per motor, it can easily be brought to light by the following approximation. Computationally, we find that $\Bfn^2 \ll \Afn\Dfn$ by a factor of thirty or more, depending on the number of flagella, which is consistent with the analysis of a single helical filament made by \citet{Purcell1997}. Hence, we neglect the $\Bfn^2$ terms and approximate the effective load per motor from Eq.~\eqref{eq:delta-load} as
\begin{equation}
    \delta \approx \frac{\Dfn\tilde{D}_b}{\tilde{D}_b+N\Dfn} 
    \label{eq:simplified-delta}
\end{equation}
where $\tilde{D}_b = D_b + NR_b^2\Cfn$ is the effective hydrodynamic resistance of rotating the spheroidal cell body and the entire $N$-filament bundle around the major axis of the cell body. The simplified Eq.~\eqref{eq:simplified-delta} reveals that the effective load per motor is a weighted average of the load on the filaments ($D_f$) and the load on the cell body ($\tilde{D}_b$), which is reminiscent of the effective resistance of multiple resistors connected in parallel. This prompts us to consider the following electrical analogy. 

\begin{figure}[ht!]
	\includegraphics[width=\textwidth]{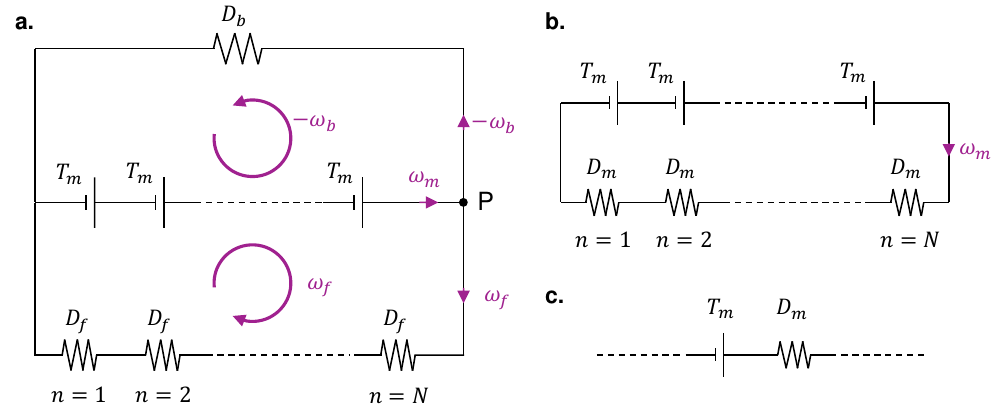}
	\caption{ {\textbf{Cargo load sharing {illustrated} via electrical analogy}. (a) In the laboratory frame, the torque balance on a swimming multiflagellated bacterium can be described as an electrical circuit with $N$ voltage sources (motors or torque-generating units). The total potential difference (torque, $NT_m$) generated by the voltage sources produces a current (angular velocity, $\omega_*$) in parallel branches of the circuit that represent the bacterial cell body and the flagellar filaments as resistors with specified electrical resistances (hydrodynamic resistance to rotation, $D_*$). The motors share the load of the cell body or `cargo'. (b) The electrical circuit in (a) can be simplified to a single-loop circuit containing $N$ voltage sources with internal resistances $D_m$ given by Ohm's law, Eq.~\eqref{eq:Ohms-law}. (c) Each individual motor perceives the combined effect of rotating the  flagellar filament and cell body in opposite directions as equivalent to rotating a single propeller with effective hydrodyhamic resistance $D_m$.}}
	\label{fig4}
\end{figure}

The hydrodynamic description of a swimming multiflagellated bacterium turns out to be analogous to the electrical circuits depicted in Fig.~\ref{fig4}. Angular velocities, torques, and hydrodynamic resistances are analogous to electrical currents, potential differences, and electrical resistances, respectively. In this framework, the multiflagellated bacterium is represented as an electrical circuit with $N$ voltage sources (motors) each supplying a potential difference (torque) of $T_m$, and generating a current flow (rotation) in two parallel branches of the circuit: the cell body represented by a resistor of resistance $D_b$, and the flagellar filaments represented by $N$ resistors of resistance $D_f$ connected in series (see Fig.~\ref{fig4}a). The closure of the electrical circuit signifies the torque balance on the independently swimming bacterium. Also, the conservation of current at node P, $\omega_m = \omega_f - \omega_b$, conveys that all components of the flagellum rotate at the same absolute speed, although we measure the angular velocity of the motor and that of the flagellar filament in different frames of reference. 

The circuit described in Fig.~\ref{fig4}a is perceived from a laboratory frame but the BFM, a complex piece of molecular machinery embedded in the cell wall, does not distinguish between different sources of resistance. The motor only feels an effective load $D_m$ which resists the generated torque or potential difference, as shown in Fig.~\ref{fig4}c. To connect the laboratory frame perspective in Fig.~\ref{fig4}a and the individual motor perspective in Fig.~\ref{fig4}c, we note that the composite circuit in Fig.~\ref{fig4}a can be simplified to a single-loop circuit as shown in Fig.~\ref{fig4}b. By applying Ohm's law and imposing that the angular velocity of the motors, $\omega_m$, is the same in the two circuits which are equivalent by construction, we determine that
\begin{equation}
    \frac{NT_m}{ND_f}+\frac{NT_m}{D_b} = \frac{NT_m}{ND_m},
    \label{eq:Ohms-law}
\end{equation}
and thus we deduce that the effective hydrodynamic load perceived by each  motor is
\begin{equation}
    D_m = \frac{D_f D_b}{D_b + N D_f}.
\end{equation}
This is equivalent to the approximate expression for $\delta$ in Eq.~\eqref{eq:simplified-delta} that was derived from the full solution of the hydrodynamic governing equations. 

What the electrical analogy illustrates intuitively is that the motors share the burden of rotating the bacterial cell body, which leads to a reduction in the effective load per motor as the number of flagella increases.  

\subsection{Critical flagellar number agrees with empirical observations and increases with viscosity}

Our theory predicts that {the swimming speed of bacteria would plateau} if the flagellar bundle contained more than a critical number of {four} filaments, as seen from the {negligible} speed increments in Fig.~\ref{fig3}d. These results correspond to bacteria with the typical cell body size and flagellar geometry of \textit{E.~coli}, swimming in the same ambient viscosity as water at 23$^{\circ}$C (Table \ref{table}). Remarkably, our estimates for the critical number of flagella are in good agreement with the flagellar numbers reported for \textit{E.~coli} bacteria, which typically have $3.3 \pm 0.9$ flagellar filaments \cite{Turner2000,Turner2016}.

The various model parameters listed in Table \ref{table} influence the critical number of flagella. Of particular biological interest is the dynamic viscosity of the environment. Bacterial motility is known to depend on both the viscosity \cite{schneider1974effect,Atsumi1996,qu2020effects} and the polymeric or colloidal micro-structure of the environment \cite{magariyama2002mathematical,patteson2015_labArratia, qu2018_labBreuer,Kamdar2022} in a motor-load dependent manner \cite{martinez2014flagellated}. Here, we focus on Newtonian fluids with varying dynamic viscosity above that of water at room temperature.

\begin{figure}[t]
	\includegraphics[width=\textwidth]{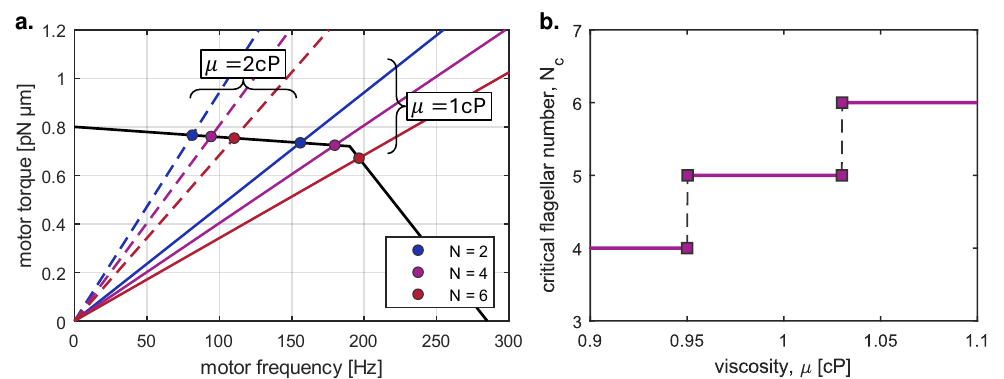}
	\caption{\textbf{Effect of increasing viscosity}. (a) The load on each motor is proportional to the ambient viscosity, thereby delaying the transition from high-load to low-load regime until a larger number of flagella in more viscous environments. (b) The critical flagellar number (i.e., the greatest integer associated with a positive speed increment) increases with viscosity.}
	\label{fig5}
\end{figure}

Because the effective viscous load on each motor is proportional to viscosity (Fig.~\ref{fig5}a), we expect the motors to operate in a high-load regime up to a larger number of flagella when bacteria swim in highly viscous environments. The delayed transition between high-load and low-load operation leads to an increase in the critical number of flagella with increasing ambient viscosity (Fig.~\ref{fig5}b). This suggests that multiflagellated bacteria could enhance their swimming capacity by growing additional flagella in more viscous media. 

Since the early days of microbial research it has been postulated that polar monotrichous bacteria are best adapted to aquatic environments where the viscosity is low (e.g., marine bacteria), whereas peritrichous flagellated bacteria are most efficient in locomotion through highly viscous environments (e.g., soil bacteria) \cite{leifson1960atlas}. Subsequent studies have confirmed that increasing the ambient viscosity can induce the production of lateral flagella in some bacterial species including \textit{Vibrio parahaemolyticus} and \textit{Bradyrhizobium diazoefficiens} \cite{Belas1986, Quelas2016}. While no direct evidence has been reported for \textit{E. coli} and other enterobacteria, the current findings will motivate further inquiry into the mechanisms and time scales over which peritrichous bacteria might adapt their flagellar number to changes in environmental viscosity.

\subsection{Critical number of flagella is determined by motor parameters}

\begin{figure}[t!]
	\includegraphics[width=\textwidth]{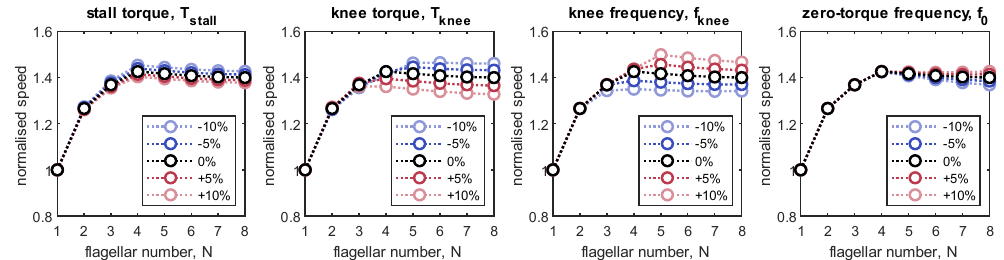}
	\caption{\textbf{Sensitivity of the swimming speed to changes in the motor parameters}. The stall torque, knee torque, knee frequency, and zero-torque frequency are varied by as much as $\pm$ 10\% around the default parameter values listed in Table \ref{table}. Each curve is normalised by its corresponding speed for a single flagellum.}
	\label{fig6}
\end{figure}

{
The BFM torque-speed relationship varies across bacterial species and is also influenced by temperature and other environmental conditions \cite{Sowa2008}. Furthermore, there is a wide range of values reported in the literature for the stall torque of bacterial flagellar motors, summarised most recently in Ref.~\cite{Das2018}. To test the applicability of our results to experimental data from other studies, we investigate the sensitivity of the swimming speed predicted by our model to the parameters describing the shape of the BFM torque-speed curve (Fig.~\ref{fig6}). We find that the swimming speed is most sensitive to changes in the knee torque and frequency, which is consistent with our understanding that bacterial slowdown occurs when the motor transitions from the high-load to the low-load regime.  Nevertheless, our theoretical predictions are robust, as changes of $\pm$ 10\% around the default parameter values lead to changes of at most 9\% in the absolute swimming speed for any given flagellar number. The results of our sensitivity analysis, shown in Fig.~\ref{fig6}, demonstrate that the critical number of flagella is determined by the location of the `knee' in the torque-speed curve relative to the fan of load lines previously depicted in Fig.~\ref{fig2}b. Recall that the slope of the load lines is determined by the viscous drag on the flagella and cell body (Eq.~\eqref{eq:delta-load}), suggesting that the critical number of flagella to achieve maximum swimming speed is controlled by both the geometry of the bacteria and the torque-generating capacity of their motors, and hence it will be specific to each bacterial species. 
}

\section{Conclusion}

In this article, we have introduced a kinematic model of bacterial swimming (Fig.~\ref{fig2}a) that resolves both the full hydrodynamic interactions (HIs) between flagellar filaments, and the torque-speed characteristics of the rotary motors actuating these filaments. Our theoretical model predicts that bacteria would slow down above a critical number of flagella (Fig.~\ref{fig2}d,e) due to a decrease in the viscous load per motor, and a subsequent reduction in the torque supplied by each motor (Fig.~\ref{fig2}b,c). We confirm that this slowdown emerges from a combination of load-dependent torque generation by the motors, and load-reducing cooperativity between the flagella (Fig.~\ref{fig3}). The cooperativity has a hydrodynamic component due to long-range interactions between the flagellar filaments, and a non-hydrodynamic component due to `cargo' load sharing between the motors. We illustrate the latter mechanism using an electrical analogy (Fig.~\ref{fig4}). {Our sensitivity analysis demonstrates that the swimming speed is susceptible to the position of the `knee' in the torque-speed curve (Fig.~\ref{fig6}). This suggests that the critical number of flagella for each bacterial species may correspond to a distinguished balance between its hydrodynamic drag signature (which is controlled by the geometry of the cell body and flagella) and the torque-generating capacity of its motors (which is notably characterised by the knee torque and frequency).}

The biological relevance of our theoretical predictions is confirmed by comparison with empirical observations. For standard swimming conditions (see Table \ref{table}), our model predicts a critical number of flagella {around four} (Fig.~\ref{fig3}d), which is remarkably close to the flagellar numbers measured for \textit{E. coli} bacteria swimming in similar conditions \cite{Turner2000,Turner2016}. Investigating the effect of ambient viscosity yields further biophysical insights (Fig.~\ref{fig5}). Our model predicts that the critical number of flagella to maximize swimming speed increases with environmental viscosity, hinting at a mechanical advantage of growing more flagella in media of higher viscosity. This finding is consistent with empirical observations \cite{Belas1986, Quelas2016} and will motivate new studies to investigate the mechanisms through which peritrichous bacteria might adapt their flagellar number to changes in environmental viscosity, and quantify the time scales for this adaptation.   

While our theoretical model offers valuable insights into bacterial multiflagellated swimming, it is essential to acknowledge its intrinsic limitations and possibilities for extension. Our model does not resolve the HIs between the flagellar filaments and the cell body, only those between the flagellar filaments. These interactions are expected to dominate the physics of bacterial swimming due to the close spacing between filaments within the flagellar bundle. Furthermore, the flagellar filaments are modelled as perfectly rigid, and there is no flexible hook joining the motor and the helical filament. A complete elastohydrodynamic model would need to account for the elasticity of both flagellar filament and hook~\cite{Nguyen2018}. Because our constant torque results are similar to those of \citet{Nguyen2018}, we do not expect the introduction of elasticity to qualitatively change the main conclusions of our study, although it may alter the theoretically predicted critical number of flagella. Finally, a significant limitation of our model is that the filaments are rotating around prescribed axes of rotation parallel to the cell body. Therefore, our model cannot capture the bundling of bacterial flagella \cite{Turner2000,Kim2003}, or the wiggling trajectories of bacteria caused by the finite angle between the cell body and the flagellar bundle \cite{Hyon2012}. 

Since the theoretical predictions are sensitive to the effective load on motors, future studies could extend our model of steady, synchronised flagellar dynamics towards an unsteady description of flagellated swimming in which the hydrodynamic load on individual filaments can vary due to changes in their relative positions (e.g., due to bundling), while the motors can respond to perceived changes in load via dynamic remodelling and stator recruitment \cite{Nirody2019, Wadhwa2022}. Although the mechanisms for controlling the location and number of flagella are not well understood \cite{Schuhmacher2015}, the average number of flagella appears to be consistent across empirical observations of the same bacterial species. To fully understand the robustness of bacterial flagellar numbers, and whether it is related to bacterial motility, future work must compare the fitness gains of multiflagellarity with the energy costs of assembling and operating the bacterial flagella \cite{Schwarz-Linek2016,Schavemaker2022}.

\begin{acknowledgments}
We are grateful to Xiang Cheng, Dipanjan Ghosh, and Shashank Kamdar for thoughtful discussions on the subject of bacterial swimming. This project was supported by the University of Cambridge (George and Lillian Schiff Studentship supporting M.T.C.), by Clare College, Cambridge (Lynden-Bell Research Fellowship supporting M.T.C.) and by the European Research Council through the European Union’s Horizon 2020 research and innovation programme (under grant agreement No. 682754 to E.L.).
\end{acknowledgments}

 \bibliography{mybibliography}

\end{document}